\documentclass[epj]{svjour}
\usepackage{amssymb}
\usepackage{graphicx}
\def\bs{\hbox{\vrule width 4pt height 4pt depth 0pt}}
\begin{document}
\title{Theoretical approach and impact of correlations on the critical packet generation rate in traffic dynamics on complex networks}
\titlerunning{Theoretical approach and impact of correlations on the traffic dynamics on complex networks}
\author{Piotr Fronczak\inst{1,}
 \thanks{e-mail: fronczak@if.pw.edu.pl}
}                     
\offprints{}          
\institute{Faculty of Physics, Warsaw University of Technology, Koszykowa 75, PL-00-662 Warsaw, Poland
}
\date{Received: date / Revised version: date}
%
\abstract{
Using the formalism of the biased random walk in random uncorrelated networks with arbitrary degree distributions, we develop theoretical approach to the critical packet generation rate in traffic based on routing strategy with local information. We explain microscopic origins of the transition from the flow to the jammed phase and discuss how the node neighbourhood topology affects the transport capacity in uncorrelated and correlated networks. } 
\maketitle
 Transport phenomena in real networked communication systems, such as the Internet \cite{Pastor2001} and WWW \cite{Albert1999}, has turned recently more and more attention in physical and computational science. Since the rapid development of society entails demands for high transport efficiency, scientists strive to develop methods for understanding and controlling traffic congestion on communication systems.

 In the basic models frequently used to mimic transport phenomena in communication networked systems \cite{Guimera2002,Zhao2005,Tadic2004,Guimera2002b,Mukherjee2005}, all nodes in a network are equally considered as hosts and routers for generating and delivering packets. Then, the traffic dynamics is defined as follows:
\begin{itemize}
  \item At each time step, there are $R$ packets generated in the system, with randomly chosen sources and destinations.
  \item During the next time steps packets travel around the network, and look for their destination-nodes. Once a packet arrives at its target, it is removed from the system.
  \item To navigate packets, each node performs a local search among its neighbors. If the packet's destination is found within the search area, it is delivered directly to the target. Otherwise, the particle is forwarded to the next node according to the prescribed strategy.
  \item At each time step every node can distribute/deliver at most $C$ packets towards their destination (the fixed value of $C$ reflects the limited router bandwidth).
  \item The queue length of each node is assumed to be unlimited and the FIFO (first in first out) discipline is applied.
\end{itemize}

In these models, one can distinguish between two kinds of strategies:

\begin{itemize}
\item In the first kind, each node has the global topological information about the network, which allows packets to be forwarded either following the shortest path \cite{Guimera2002,Zhao2005,Zhao2004,Holme2002} or using the concept of betweenness \cite{Guimera2002b,Barthelemy2003,Huang2009}, which measures the number of total shortest paths that pass through the given node. This kind of strategy may be practical for small or medium size networks, but not for very large networks in real communication systems such as the Internet, WWW, peer-to-peer networks \cite{Wang2006} or urban transportation systems \cite{Hu2008,Scellato2010}.
\item The strategies of the second kind base on local information (each node only knows its neighbourhood)and are favored in very large networks due to heavy communication cost of searching.
\end{itemize}

One of the most important measurements for transport performance of a network is the traffic capacity, $R_c$, i.e. the critical packet generation rate. At $R_c$, the network undergoes a phase transition from the free flow state to the congested state. When the packet
generation rate $R$ is below $R_c$, the number of generated and delivered packets are balanced and therefore the network is in free flow state. On the other hand, when $R$ goes beyond $R_c$, the number of packets keeps on increasing with time and leads to congestion,
simply because nodes cannot deliver too many packets at each time step due to limited delivering capacity.

 Although a number of empirical strategies for improving transport efficiency has been proposed (see the review \cite{Chen2011} and references therein, and also \cite{Martino2009}), the theoretical background of traffic congestion phenomena is not well developed. With reference to this theoretical line of research we would like to highlight Ref. \cite{Zhao2005}, where the estimation of traffic capacity for the stategies of the first kind has been provided.

In this paper, we present a theoretical approach to the critical packet generation rate applied to the strategy of the second kind proposed by Wang et al. \cite{Wang2006b}. The strategy is based on the biased random walk. In this strategy the next position of the packet (node $j$) is chosen according to the prescribed preferential transition probability $w_{ij}$
\begin{equation}\label{wlj}
w_{ij}=\frac{k_j^{\alpha}}{\sum_{m=1}^{k_i}k_m^\alpha},
\end{equation}
where the sum in the denominator runs over neighbors of the node $i$, which represents the current position of the packet, and the exponent $\alpha$ is the model free parameter.
Note that according to the formula (\ref{wlj}) the transition
probability from $i$ to $j$ depends only
on the connectivity of the next-step node $j$. Note also that for
$\alpha=0$ we recover the ordinary unbiased random walk studied by
Noh and Rieger \cite{Noh2004}.
In the model, the so-called path iteration avoidance is assumed, which means that no link can be visited twice by the same packet.

The main result of this model is presented in Fig. 1. The black squares (numerical simulations) show that the optimal performance of the system (the largest traffic capacity) corresponds to $\alpha=-1$, which represents the anti-preferential transition probability
$w_{ij}\sim 1/k_j$. An interesting finding arises from the comparison of Fig. 1a and 1b showing results for classical random graphs and networks with power law node degree distribution, respectively. In the former case, the function $R_c(\alpha)$ has a smooth shape. In the later case the character of the relationship changes sharply. In what follows, we explain the observations with the help of a simple theoretical approach.

\begin{figure}
\includegraphics[width=7.8cm]{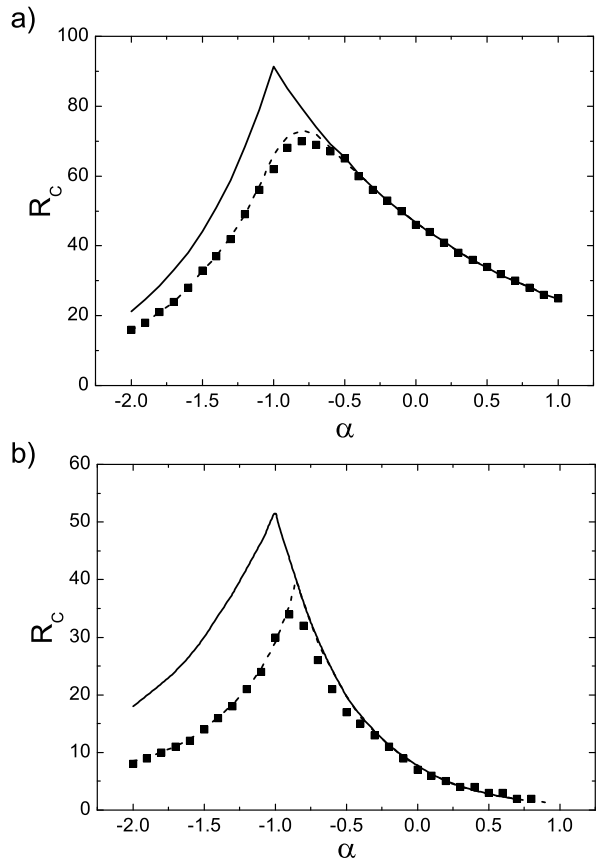}
\caption{The critical packet generation rate $R_c$ calculated for different values of the parameter $\alpha$ with network size $N=1000$ and capacity $C=10$ in a) classical random graphs and b) scale-free networks $P(k)\sim k^{-\gamma}$ with the characteristic exponent $\gamma=3$. Relatively large value of average node degree ($\langle k\rangle=12$ in case (a) and $\langle k\rangle=9.3$ in case (b)) ensures connectivity of the networks, i.e. there exists a path between each pair of nodes. Black squares represent results of numerical simulations, solid lines correspond to theoretical prediction of Eq. (7), while dashed lines have been calculated from Eq. (7) with $P_i^{\infty}$ replaced by $P_{i\vartriangle} ^{\infty}$.}
\label{fig1}
\end{figure}

Let us start with the simple observation: the congested phase occurs when the average number of packets $N_p(i,t,R)$ arriving at a certain node $i$ at time $t$ exceed its processing capacity $C_i$. Thus the critical value of the posting rate $R_c$ should be somehow found form the balance condition
\begin{equation}
C_i=N_p(i,t,R_c),
\end{equation}
which, for $C_i=const$, can be also rewritten as
\begin{equation}\label{C}
C=N_p(i,t,R_c).
\end{equation}
Note that in the free flow state one has $C>N_p(i,t,R)$ whereas in the congested phase there is $C<N_p(i,t,R)$. In the considered model, $N_p(i,t,R)$ can be written as
\begin{equation}\label{Npi}
N_p(i,R)=P_i^{\infty} N_p(R),
\end{equation}
where the stationary occupation probability, $P_i^{\infty}$, describes the probability that the particle is located at the node $i$ in the infinite time limit (for that reason we have ommited time dependence of $N_p$) and $N_p$ is the total number of packets in the network. It has been shown \cite{Fronczak2009}, with the help of biased random walk formalism, that
\begin{equation}\label{Pinfty}
P_i^\infty=\frac{k_i^{\alpha+1}}{N\langle k^{\alpha+1}\rangle},
\end{equation}
where $N$ is the network size. Note that for $\alpha=0$, which stands for the unbiased random
walk, the stationary distribution is, up to normalization, equal
to the degree of the the node $i$, i.e. $P_i^\infty\sim k_i$. It
means that the more links a node has, the more often it will be
visited by a random walker. Note also that for $\alpha=-1$, the stationary occupation probability is
uniform $P_i^{\infty}=1/N$. The same scaling behavior as given by Eq. (\ref{Pinfty}) was found in Ref. \cite{Wang2006b} for the number of packets moving simultaneously on BA networks \cite{Barabasi1999} in the free flow state. It means that in the free flow, and also in the critical point which is the limiting case of free flow state, the packets may be considered as non-interacting particles (i.e. independent biased random walkers).

The last observation allows us to calculate the total number of packets distributed over the whole network $N_p$ in the free flow state as:
\begin{equation}\label{Np}
N_p=R\langle T_{ij}\rangle,
\end{equation}
where $\langle T_{ij}\rangle$ stands for the mean first-passage time averaged over all pairs of nodes. $\langle T_{ij}\rangle$ can be understood as the mean lifetime of a packet and can be calculated theoretically from Eq. (19) in \cite{Fronczak2009}.

Now, combining Eqs. (\ref{C}), (\ref{Npi}) and (\ref{Np}) one can find the critical value of packets generation rate $R_c$:
\begin{equation}\label{Rc}
R_c=\frac{C}{P_i^{\infty} \langle T_{ij}\rangle}.
\end{equation}
The last equation shows that the critical value of packets generation rate is, due to $P_i^{\infty}$, a function of a node degree $k_i$. It means that in heterogeneous networks  nodes of different degrees become congested for different values of $R$. It also means that although the system as a whole enters the jammed state even if one node is congested, it still possesses partial capacity for forwarding packets in this phase.

To find the critical value of $R_c$, which reflects simulation results presented in Fig. 1, from the whole set of different degree dependent values of $R_c$ one has to choose the smallest one. Since $R_c$ is inversely proportional to $P_i^{\infty}$, choosing minimal value of $R_c$ corresponds to taking nodes with the highest $P_i^{\infty}$. In Fig. 2, $P_i^{\infty}$ is presented as a function of a node degree $k$ for different values of the parameter $\alpha$ and for two different network topologies. Solid lines correspond to theoretical prediction of Eq. (\ref{Pinfty}) whereas black squares are results of numerical simulations. In both kinds of networks for $\alpha<-1$, $P_i^{\infty}$ is maximal for nodes with the smallest degrees and in this range of the parameter $\alpha$ these nodes become congested first. For $\alpha>-1$ the situation changes and congestion starts in nodes with the highest degeees.

\begin{figure}
\includegraphics[width=8.4cm]{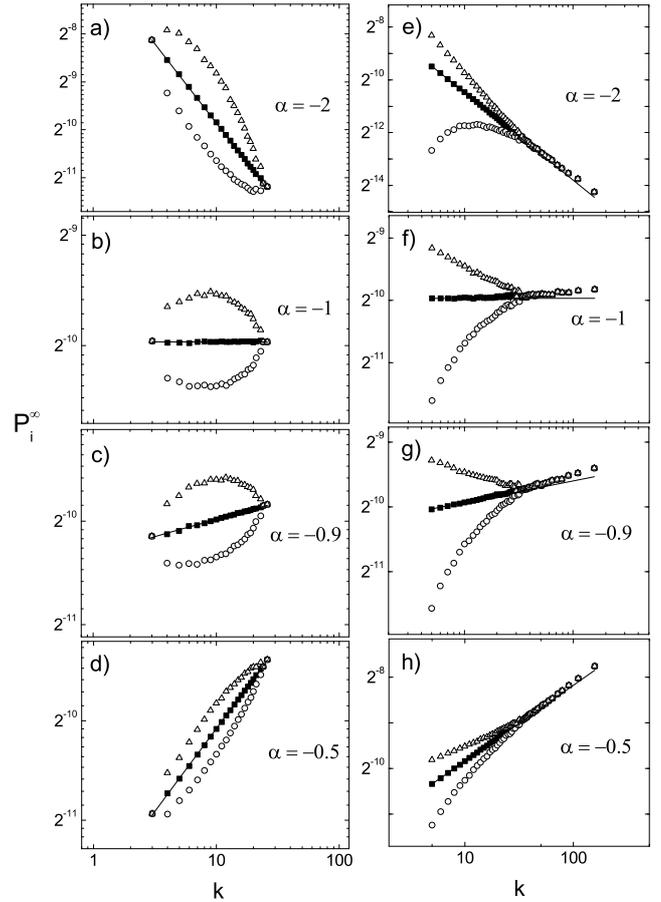}
\caption{Stationary occupation probabilities calculated for different node degrees $k$ and different values of parameter $\alpha$ in classical random graphs (a)-(d) and scale free networks (e)-(h). Solid lines correspond to theoretical prediction of Eq. (\ref{Pinfty}). Black squares, triangles and circles are the results of numerical simulations of $P_{i\bs}^{\infty}(k)$, $P_{i }^{\infty}(k)$ and $P_{i \circ}^{\infty}(k)$, respectively.}
\label{fig2}
\end{figure}

Solid lines shown in Fig. 1 represent theoretical estimation of $R_c(\alpha)$ as given by Eq. (\ref{Rc}). The lines have been calculated taking into account $P_i^{\infty}(k_i=k_{min})$ for $\alpha<-1$ and $P_i^{\infty}(k_i=k_{max})$ for $\alpha>-1$. The mean lifetime of a packet $\langle T_{ij}\rangle$ in Eq. (\ref{Rc}) has been calculated numerically because its theoretical estimation given by averaging Eq. (19) in Ref. \cite{Fronczak2009} due to applied approximation (see Eq. (16) in Ref. \cite{Fronczak2009}) gives correctly only the order of magnitude of $\langle T_{ij}\rangle$.

Although the character of theoretical lines in Fig. 1 reflects the shape of numerical results and confirms the numerically found optimal performance of the system for $\alpha = -1$, it is still far from perfection. The discrepancies are caused by the fact, that Eq. (\ref{Pinfty}) describes only the average occupation of a node with a given degree $k$. In reality, $P_i^{\infty}(k)$ may differ among the nodes of the same degree. In general, the differences depend on the topology of a node neighbourhood. Therefore one can write formally
 \begin{equation}\label{Psquare}
P_i^{\infty}(k)\equiv P_{i \bs}^{\infty}(k)=\langle p_{i,\Gamma}^{\infty}(k)\rangle,
\end{equation}
 where $p_{i,\Gamma}^{\infty}(k)$ is a stationary occupation probability for a node with particular neighbourhood topology $\Gamma$ and the average is calculated over all possible such topologies. The black square symbol in Eq. (\ref{Psquare}) has been used to reflect the fact that we are talking about results marked by black squares in Fig. 2.

 The triangles(circles) in Fig. 2 represent stationary occupation probability calculated for the most(the least) frequently occupated node among all nodes with the same degree (the results are averaged over $100$ network realizations to get rid of fluctuations)

\begin{eqnarray}\label{Ptriangle}
P_{i \vartriangle}^{\infty}(k)=\max_\Gamma \{ p_i^{\infty}(k)\},\nonumber\\
P_{i \circ}^{\infty}(k)=\min_\Gamma \{ p_i^{\infty}(k)\}.
\end{eqnarray}

The explained above scenario of congestion means that the nodes which become congested first are those represented by triangles rather than by black squares. This observation allows to understand the difference in shape of $R_c(\alpha)$ between classical random graphs and networks with power law node degree distribution (cf. Fig. 1). In the former case, along with the increase of the parameter $\alpha$ the initial congestion affects nodes with gradually larger and larger degrees. In the later case of scale free networks large negative values of $\alpha$ correspond to congrestion of the nodes with the lowest degrees. Nearby $\alpha \approx -0.9$ (not $\alpha=-1$ as was estimated in previous studies \cite{Wang2006b}) situation changes suddenly and the highly connected nodes become responsible for the congestion (without the intermediate participation of nodes with middle degrees). The dashed line in Fig. 1 calculated from Eq. (\ref{Rc}) with $P_i^{\infty}$ replaced by the discussed above maximal stationary occupation probability shows excellent agreement with numerical simulations.

As we have stated before, stationary occupation probability $P_i^{\infty}$ calculated for nodes of the same degree is in fact the average over all possible topologies of a node neighbourhood (cf. Eq. (\ref{Psquare})). To show the impact of neighbourhood topology on the probability we have calculated the average degree of the nearest neighbor, $k_{nn}$, for three groups of nodes: those marked in the Fig. 2 by squares (all the nodes), triangles (the most frequently occupated nodes) and circles (the least frequently occupated nodes). The results for SF networks and for $\alpha=-1$ (corresponding to the scenario of unbiased random walk shown in Fig. 2f) are presented in Fig. 3a. Although, in the case of uncorrelated networks, $k_{nn}$ calculated for all nodes does not depend on $k$ (what is confirmed by horizontal character of black squares), the Fig. 3a shows that the neighbourhood of the nodes which are responsible for congestion (i.e. in case of $\alpha=-1$ the most frequently occupated nodes with $k=k_{min}$) is composed of weakly connected nodes. Then one can suggest that increasing their $k_{nn}$ (i.e. making their neighbourhood disassortatively correlated) may improve transport capacity of the network.

\begin{figure}
\includegraphics[width=8.4cm]{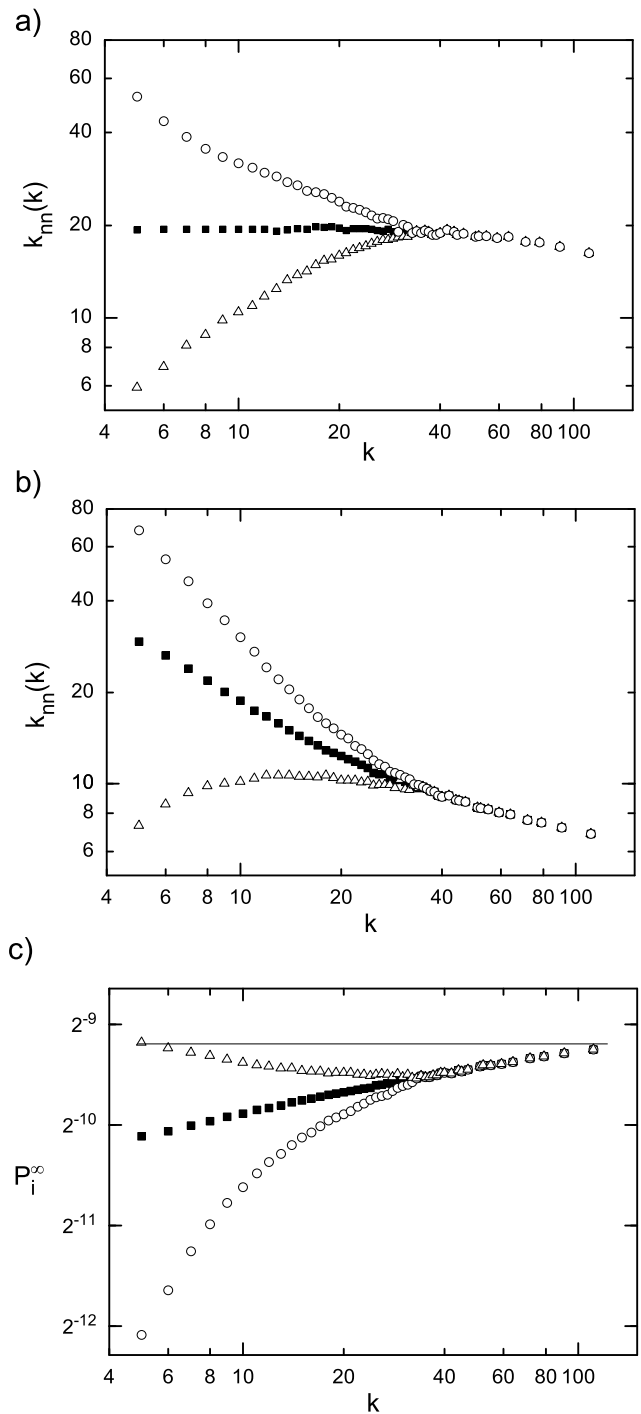}
\caption{The average degree of the nearest neighbor for all the nodes (squares), the most frequently occupated nodes (triangles) and the least frequently occupated nodes (circles) for $r=0.00$ (a) and $r=-0.20$ (b). Stationary occupation probabilities calculated for different node degrees $k$ for $\alpha=-1$ in SF networks (c).}
\label{fig3}
\end{figure}

To check the hypothesis one has to introduce degree correlations to the network. From many methods for generating correlated networks (e.g. \cite{Boguna2003,PF2}) we select one of the simplest model proposed by Noh \cite{Noh2007,ERG}. The model belongs to a class of the exponential random graph family \cite{Newman2004,PF1}. In this class, a network model is defined as a Gibbsian ensemble of
networks with an associated network Hamiltonian
\begin{equation}\label{JHamiltonian}
H(G) = -\frac{J}{2} \sum_{i,j=1}^N a_{ij} k_i k_j \ ,
\end{equation}
where $J$ is a control parameter and $a_{ij}$, element of the adjacency matrix, takes the value $a_{ij}=1 \mbox{\ or\ } 0$ if nodes $i$ and $j$ are connected or not. A positive~(negative) correlation is favored by a positive~(negative) value of $J$.
The Monte Carlo dynamics is based on updating network configurations via the link rewiring process, which preserves the degree of each node. The dynamics leads to the Gibbsian ensemble in the stationary state.

The assortativity of the network \cite{ERG} is measured by the Pearson correlation coefficient of the degrees at either ends of an edge:
\begin{equation}\label{assortativity}
r = \frac{ \langle kk'\rangle_{l} - \langle(k+k')/2\rangle_{l}^2}
    { \langle( k^2 + k'^2)/2\rangle_{l} - \langle (k+k')/2\rangle_{l}^2} \ ,
\end{equation}
where $\langle \cdot\rangle_{l}$ denotes the average over all links, whereas $k$ and $k'$ represent the degrees of two nodes at either end of links. The sign of $r$ indicates a positive~(assortative) or negative~(disassortative) degree correlation. It vanishes for uncorrelated networks.

The stationary state value of the assortativity for SF network is presented in Fig. 4. The possible values of $r$ in correlated SF networks, which can be obtained with the help of thi model belong to the range $-0.27\leq r\leq 0.27$. This differentiates SF networks from networks with the Poisson degree distribution for which $-1\leq r\leq 1$ (cf. Fig. 2b in \cite{Noh2007}). It seems that the possible patterns of correlation in scale-free networks are restricted by the power law degree distribution.

\begin{figure}
\includegraphics[width=8.4cm]{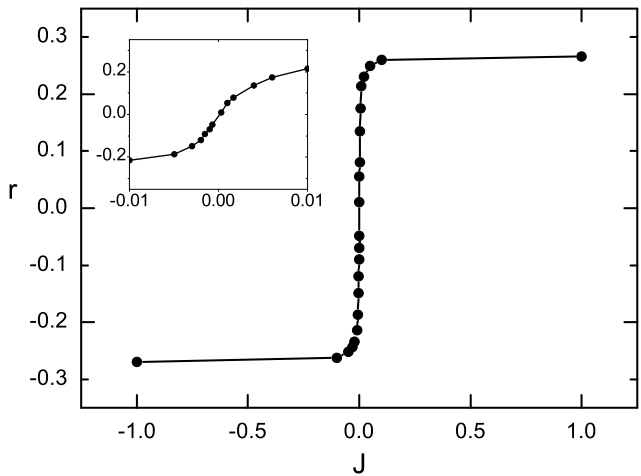}
\caption{Stationary state values of the assortativity as a function of $J$ in SF network.}
\label{fig4}
\end{figure}

Having the tool for generating correlated networks in hands, one can repeat the calculations of $k_{nn}$ for such networks. In Fig. 3b we have shown results for the case of SF networks with  $r=-0.20$. In such networks $k_{nn}$ for the nodes responsible for congestion has increased from about $5.9$ (in uncorrelated networks) to about $7.3$. Unfortunately, further increasing of negative correlations will have an undesirable effect on the network capacity: detaching weakly connected nodes from other weakly connected nodes one have to attach them to the highly connected ones. As a consequence, the transport abbilities of hubs decrease. In the Fig. 3c we have shown corresponding occupation probabilities for that case. As one can see, $P_i^{\infty}$ for the most frequently occupated nodes with $k_{min}$ and $k_{max}$ equalize (what is emphasized by the horizontal line). Therefore, the case of $r=-0.20$ is optimal, because any change of network correlations can only increase the maximal occupation probability (through increasing $P_{i \vartriangle}^{\infty}(k_{min})$ or $P_{i \vartriangle}^{\infty}(k_{max})$).

Finally, in the Fig. 5 we have shown the profiles of the critical packet generation rate $R_c$ calculated for three different values of the parameter $r$: in the case of uncorrelated network ($r=0.00$), the case of optimally correlated network ($r=-0.20$), and the case of highly correlated network ($r=-0.27$). The interesting observation is that although in the most optimal case, $k_{nn}$ calculated for the nodes responsible for congestion has increased just about one degree (cf. Fig. 3b), in the same time the critical packet generation rate $R_c(\alpha=-1)$ has increased  from $29$ to $41$, i.e. about $40\%$.

\begin{figure}
\includegraphics[width=8.4cm]{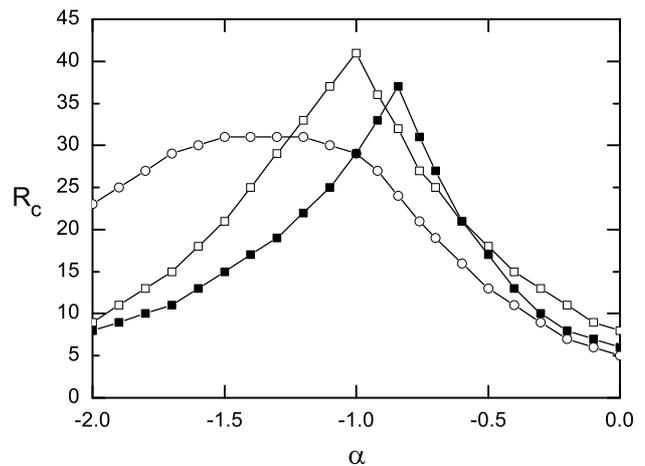}
\caption{Profiles of the critical packet generation rate $R_c$ calculated for the case of uncorrelated network ($r=0.00$) (black squares), the case of optimally correlated network ($r=-0.20$) (open squares), and the case of highly correlated network ($r=-0.27$) (open circles).}
\label{fig5}
\end{figure}

The observation, that transport capacity of a network can be enhanced for both assortative mixing and disassortative mixing has been recently reported in Ref. \cite{Sun2009}. In the metioned paper, authors have studied routing strategies with the global topological information. In our case, assorative mixing cannot  improve the network capacity, since it only decrease $k_{nn}$ of nodes responsible for congestion.

In summary, using the formalism of the biased random walk in random uncorrelated networks with arbitrary degree distributions, we have developed the theoretical approach to the critical packet generation rate in traffic dynamics with the local routing strategy as proposed by Wang et al. We have shown that the random walk approach can be used to give microscopic explanation of the phase transition from
free flow to the jammed phase. We have also discussed the effect of degree correlations and node neighbourhood topology on the properties of transport in complex networks.

\section*{Acknowledgements}
The author wishes to thank Dr. Agata Fronczak for
her valuable comments and suggestions.
This work was financially supported by internal funds of the Faculty of Physics at Warsaw University of Technology.


\end{document}